
\magnification=1200
\baselineskip=0.460truecm
\centerline{\bf FOUR DIMENSIONAL CONFORMAL GRAVITY,}
\smallskip
\centerline{\bf CONFINEMENT, AND GALACTIC ROTATION CURVES}
\vskip 0.30truecm
\centerline {\bf Philip D. Mannheim}
 \vskip 0.10truecm
\centerline {Department of Physics, University of Connecticut, Storrs, CT
06269-3046}
\vskip 0.10truecm
\centerline{(mannheim@uconnvm.uconn.edu)}
\vskip 0.15truecm
\centerline{preprint UCONN-94-3, July 1994, to appear in the proceedings of
PASCOS 94}
\vskip 0.30truecm
\noindent
{\bf Abstract}$~~~$ We review the
current status and prospects for the conformal invariant fourth order theory of
gravity
which has recently been advanced by Mannheim and Kazanas$^{1-9}$ as a candidate
alternative
to the standard second order Einstein theory. We examine how it is possible in
principle to
replace the Einstein theory at all while retaining its tested features, and we
appeal to the
wisdom gleaned from particle physics to suggest a covariant alternative.
Specifically,
motivated by the underlying scale invariance of the fundamental strong,
electromagnetic and
weak interactions, we propose that, just like their inertial counterparts in
the other
fundamental theories, gravitational mass and length scales should also be
induced dynamically
by spontaneous breakdown in a theory of gravity which itself possesses no
fundamental scales
at all. Indeed, in a sense  this viewpoint is even mandated by the equivalence
principle since  once inertial masses are dynamical, it is natural to expect
gravitational
masses to be dynamical too.  Given the connection of conformal gravity to
particle theory,
we explore the theory as a microscopic fundamental theory of elementary
particles, to thus
give curvature a more prominent role in elementary particle dynamics, and in
particular in
the generation both of elementary particle masses  and the extended bag-like
models which are
thought to describe them. We discuss the degree to which conformal gravity can
compete with
string theory as a consistent candidate microscopic theory of gravity.
Additionally, we
explore the theory as a macroscopic theory of gravity where the exact,
non-perturbative,  classical potential is found to be of the confining
$V(r)=-\beta /r+\gamma r/2$ form, a form which enables us to provide an
explanation
for the general systematics of galactic rotational velocity curves
without the need to assume the existence of copious amounts of dark matter;
with
this explanation of the curves apparently not being in conflict with the
current round
of microlensing observations.
\bigskip
\noindent
{\bf (1) The Case for Reconsidering the Standard Einstein Theory}
\medskip
At the present time there is little doubt in the general community as to the
correctness of the standard second order Einstein theory of gravity which is
based on the action
$$I_{EH}=-\int d^4x (-g)^{1/2}R^{\alpha}_{\phantom{\alpha} \alpha}/16 \pi G
\eqno(1)$$
and gravitational equations of motion of the form
$$R_{\mu \nu} - g_{\mu \nu} R^{\alpha}_{\phantom {\alpha} \alpha}/2
= -8 \pi G T_{\mu \nu} \eqno(2) $$
However given the fact that this theory when applied to currently available
astrophysical and cosmological data then requires that
the universe be composed
of overwhelming amounts of non-luminous or dark matter, it is a well
established scientific tradition to pause and question so startling an
implication, and to at
least consider the possibility that this need for dark matter might instead
actually be
signaling a possible  breakdown of
the standard Newton-Einstein theory on the largest distance scales. Given this
possibility, there is then some value in going over the standard theory step by
step to see whether there is anything in the entire package which still has a
chance to go wrong. Thus we seek to identify
which aspects of the theory are well observationally established and which are
less so, so that we can then determine what it is that available observational
data actually mandate, with this information then to be used as a guide to
identify potential candidate gravitational theories.
We thus argue not that the standard theory might eventually
fail, but rather, we find$^{8,9}$ it to
still have a few weak links and loose ends which leave it in an exposed
position. Moreover,
identifying such loose ends is of interest in and of itself, independent of the
merits or
otherwise of the conformal theory we consider here, since their very
elucidation sharpens our understanding and appreciation of the standard theory.

While the equivalence principle is one of the key underpinnings of General
Relativity, it is important to note that the equivalence principle itself does
not address the question of what specific form the gravitational equations of
motion should take. Rather its success implies only that gravity in fact be
a metric theory, with the gravitational field to be described uniquely by a
covariantly coupled metric. Since geodesic motion follows from the covariant
conservation of the standard test particle energy-momentum tensor, and since
this local conservation itself follows purely from covariance in any general
coordinate invariant gravitational theory (something which was pointed out by
Eddington as long ago as the very early days of Relativity), one can conclude
that geodesic motion itself only entails that the gravitational action be a
general coordinate scalar. The dependence of the geodesics on the detailed
dynamics of the theory actually only comes through the fact that it is
the gravitational field equations which then determine the explicit
form of the Christoffel symbols which enter into the
geodesic equations. Moreover, Eddington also noted that since the three classic
tests of General Relativity only probe the Schwarzschild solution associated
with
the Ricci flat vacuum outside of a source, the three tests can also be
satisfied
by having some higher derivative of the Ricci tensor vanish instead (this of
course being the case in higher order gravity), since the Schwarzschild
solution
is still an exact solution in such cases. Hence by studying the region
exterior to a source no information is provided on what $T_{\mu \nu}$ might
be equal to in the interior region where it is non-zero. Or, in other words,
Einstein gravity is sufficient to explain the classic tests but not necessary.

As regards the actual structure of the equations of motion in the $T_{\mu \nu}$
non-zero region, it is curious to note that Einstein did not in fact derive the
familiar second order equations of motion of Eq. (2) by appealing to some
fundamental principle (which would incidentally be more in keeping with the
philosophical route he followed to reach the equivalence principle); rather he
simply postulated Eq. (2) by noting, first, that these equations reduce for
weak gravity
to the standard second order gravitational Poisson equation
$$\nabla^2 \phi ({\bf r}) = g({\bf r}) \eqno (3)$$
with its familiar exterior Newtonian potential solution, viz.
$$\phi(r>R) = -{1 \over r} \int_{0}^{R} dr^{\prime} g(r^{\prime})
{r^{\prime}}^2  \eqno(4)$$
for a spherically symmetric, static source with radius $R$; and that, second,
Eq. (2)  yields relativistic corrections to this non-relativistic theory. The
observational confirmation of these corrections on terrestrial to solar system
distance scales not only established the validity of the Einstein theory on
those scales but seems to have established it on all others too, even though
many other theories could potentially have the same leading perturbative
structure
on a given distance scale while differing radically elsewhere. We note the one
way nature of the argument. Second order Einstein implies second order Poisson
which in turn implies Newton. The reverse however is not true, and at the
present time the Einstein theory is only sufficient to yield Newton's Law of
Gravity, but not yet necessary, to thus
open the door to candidate alternative theories of gravity even at this late
date.

To explore this point further, consider instead a fourth order Poisson equation
$$\nabla^4 B(r)  = f(r). \eqno(5)$$
For a spherical source Eq. (5) can be completely integrated in a closed
form$^{8}$ to yield
$$B(r>R) =- {r \over 2} \int_{0}^{R} dr^{\prime} f(r^{\prime})
{r^{\prime}}^2
 - {1 \over 6r} \int_{0}^{R} dr^{\prime} f(r^{\prime}) {r^{\prime}}^4
\eqno(6)$$
as its exact
exterior solution. The fourth order Poisson equation (which incidentally
emerges as an exact
equation of the relativistic fourth order conformal gravity theory below) thus
also contains
the Newtonian potential in its solution, thus divorcing Newton's Law of Gravity
(in principle
at least) from both the second order Poisson equation and the second order
Einstein
theory, to thus make it manifest that the Einstein theory is in fact only
sufficient
to yield Newton but not necessary.
Since the fourth order Poisson equation also yields the linear potential
term in Eq. (6) which would then dominate over Newton at large
distances, our very ability to use pure Newtonian gravity
on the largest distance scales comes from the assertion
that the Newtonian term is the leading term at infinity, something which,
while widely believed, is not definitively supported by any known data at
the present time. Indeed, there is not yet any evidence for the validity of
Newton's Law of Gravity at all on galactic or larger distance scales as the
whole dark matter issue makes apparent; in fact, if galactic rotation curve
data were the only data we had (i.e. if we had no solar system information at
all),
we would not be able to extract out a Newtonian Law at all. Thus the validity
of the Einstein Equations (as opposed to the validity of covariance itself)
rests not so much on the relativistic corrections, but
rather on the validity of the second order Poisson equation and its Newtonian
solution on distance scales much larger than those on which the Newton-Poisson
picture was first established; and their validity in turn requires the universe
to be predominantly non-luminous.

Searches to actually ascertain the baryonic dark matter
content of our galaxy are currently under way via the MACHO, EROS and OGLE
microlensing observations (see the contribution of D. Bennett, PASCOS 94
proceedings).  However, the current microlensing counting rates seem to be
indicating
that while there are in fact more faint low mass stars and brown dwarfs in the
Milky Way than
we had anticipated, there does not appear to be the amount needed for rotation
curve
systematics (comment by D. N. Spergel in his talk at the PASCOS 94 conference),
and perhaps
not even for galactic stability, with a mass density in lenses perhaps of the
order of that in
luminous matter in the galaxy, rather than the sought after factor of ten or so
more.
Moreover, if a brown dwarf spherical halo is to explain rotation curves, it
must additionally
be found to be distributed with a core radius which is a factor or so larger
than the
scale length $R_0$ of the observed surface brightness ($\sim$ exp$(-R/R_0)$) of
the luminous
disk, something which is simply not known at the present time. In the fitting
to rotation
curves, a pure Newtonian exponential disk gives a peak in the rotation curve at
around
$2R_0$ (to thus require something additional at larger radii, either a
halo with a bigger core radius or some new gravity which acts on these larger
distances).
Moreover, the Newtonian disk $M/L$ mass to light ratio (taken to be a constant
for a given
galaxy) is usually normalized in the standard dark matter fits (and also in the
conformal
gravity fits shown in	Fig. (1)) to the inner region maximum in the rotation
curve. In these
so called maximum disk fits the needed $M/L$ ratio is usually found to be a
factor
larger than that actually detected optically in the solar neighborhood. Since
most of the
microlensing events that have so far been detected are due to lensing off the
bulge of the
galaxy, they are essentially confirming that there are many non-luminous
sources in the
plane of galaxy, to thus actually lend support to the maximum disk fitting
commonly used. For
the standard dark matter fits, less than maximum disk is also permitted, with
the halo (whose
distribution is totally unrelated to that of the disk) then being parametrized
so it can
contribute in the inner region as well as the outer. However, for the conformal
gravity fits
to be presented below, the new linear potential term is integrated over the
same
disk distribution as the Newtonian one and thus must be maximum disk or else
the fit would
fall below the data at all radii. Lensing off the bulge thus appears to be
supporting
the maximum disk fitting required of the conformal theory. At regards lensing
in the
direction of the Magellanic clouds which explores the spherical halo, it
currently appears
that there are just not enough events, and not even as many as off the bulge.
This might
actually prove to be a severe problem for the standard dark matter theory,
since as the known
amount of matter in the plane of the disk is found to increase, the spherical
halo would have
to contain all that much more matter again if it is indeed going to stabilize
the now much
heavier disk. (This is not a problem for the conformal theory though - D. M.
Christodoulou
(Ap. J. {\bf 372}, 471 (1991)) has shown that disks are actually stable under
conformal
gravity potentials without the need for any spherical halo at all).  While the
microlensing
counting rates will be better known in the near future, an ultimate shortfall
in the number
of halo lenses would mean either that the bulk of galactic dark matter is
non-baryonic, or
that a new gravity theory is required. Moreover, even if the requisite dark
matter is all
there galactically (which incidentally  would not actually exclude the
conformal theory but
rather simply set a strict upper bound on the strength of the linear potential
term in Eq.
(6)), and even if the conformal theory should eventually fail observationally,
that would
still only leave the Einstein theory as a sufficient theory of gravity, with
the challenge
then being to find a new fundamental  principle which would make it necessary
also. (Indeed,
it is the very absence of any such underlying fundamental principle which has
engendered
problems such as the notorious cosmological constant problem, a problem for
which the
standard theory has no current answer and simply ignores by fiat in Eq. (2)).
Since no
fundamental principle is currently forthcoming, it is thus valid to explore
alternate
covariant gravitational theories which do possess some such principle.
\medskip
\noindent
{\bf (2) Conformal Gravity as a Macroscopic Theory}
\medskip
In recent times it has become fashionable in particle physics to consider
theories with local
invariances and with no fundamental length scales at the level of the
Lagrangian at all; and indeed today all of the other fundamental interactions
(the strong, electromagnetic, and the weak interactions) are all thought to be
local scaleless gauge theories which can only acquire mass or length
scales through dynamics. Consequently, it is both attractive and a possible
road to a
unification of all the fundamental forces to
entertain the idea that gravity should also be a theory with no fundamental
length scale either, and that it should also obey some analogous
local principle; and indeed the principle of local conformal
invariance of the spacetime geometry (i.e. invariance under local
stretchings of the form $g_{\mu \nu} (x) \rightarrow$ exp$(2\alpha (x))g_{\mu
\nu}(x)$) will precisely serve this purpose since it forces upon us a
theory of gravity with no fundamental length scale at all. This
theory of gravity is known as conformal or Weyl gravity, and it is
the unique four dimensional theory of gravity which meets all of our above
stated
needs. Thus we propose that gravity be based not on the Einstein-Hilbert action
but rather on the conformal invariant fourth order action
$$I_W = -\alpha \int d^4x (-g)^{1/2}
C_{\lambda\mu\nu\kappa}C^{\lambda\mu \nu\kappa} \eqno(7)$$
where $C_{\lambda\mu\nu\kappa}$ is the conformal Weyl tensor and $\alpha$ is a
purely dimensionless coefficient. Over the years many previous authors have
considered
gravity based on Eq. (7) (Refs. (1-9) and the reviews of S. L. Adler
(Revs. Mod. Phys. {\bf54}, 729 (1982)) and of A. Zee (Ann. Phys. {\bf 151}, 431
(1983)) give
some of the bibliography), and the theory has had an up and down history, being
pronounced
dead many times but never actually being completely buried. Conformal
invariance was initially
introduced by Weyl who wanted to use the same $\alpha(x)$ for both the
conformal
transformations and the electromagnetic gauge transformations as a way to unify
the two long
range forces. Standing in the way of such a proposal was the immediate
realization that
conformal invariance implied that all particles had to be massless, and so the
theory could
not progress. However, with the advent of modern spontaneously broken gauge
theories, it is
now apparent that mass can still be generated in the vacuum in otherwise
dimensionless
theories, thereby permitting us to revisit the issue. Motivated by the way the
Fermi constant
is generated in spontaneously broken weak interactions, it is immediately
suggested to try to
dynamically induce the Einstein action with its Newtonian constant as the
macroscopic low
energy limit of a microscopic conformal theory, a program considered recently
by Adler
and by Zee. Such a program
if successful would then be sufficient to yield the standard gravitational
phenomenology (and
then of course require dark matter). However, as we noted above, such a program
is not
necessary. Observation does not require the recovering of the Einstein
Equations, only the
recovering of their solutions in the kinematic region where the solutions have
been tested.
Thus  Mannheim and Kazanas suggested to eschew looking for an effective low
energy
Einstein limit altogether, and instead to consider the fourth order
theory in and of itself as a macroscopic gravitational theory in its own right,
to try to
find exact solutions to it, and to confront it with observation directly.
(In passing we note that string theory, the current most popular microscopic
gravitational
theory, involves not only the second order Einstein term but all other orders
too. Moreover,
fourth order terms are even induced as radiative corrections in the
non-renormalizable second
order Einstein theory. Thus there is no dispute that fourth order terms play
some role in
physics -  and our interest here is simply in using data directly to determine
the relative
strengths of specific terms).

Since conformal gravity  possesses no intrinsic fundamental
length scale (thereby immediately excluding any cosmological term and
thus naturally addressing this longstanding open problem$^{1}$),
its gravitational equations must thus be fourth order rather than
second  order ones with Eq. (2) then being replaced
by$^{2}$
$$ 4 \alpha
W_{\mu \nu} =4\alpha (W^{(2)}_{\mu\nu} - W^{(1)}_{\mu\nu}/3)=T_{\mu\nu}
\eqno(8)$$
where
$W_{\mu  \nu}$ is given by
$$W^{(1)}_{\mu\nu} = 2 g_{\mu\nu}(R^{\alpha}_{\phantom{\alpha}\alpha})
^{;\beta}_{\phantom{;\beta};\beta}
-
2(R^{\alpha}_{\phantom{\alpha}\alpha})_{;\mu;\nu}
-2 R^{\alpha}_{\phantom{\alpha}\alpha}
R_{\mu\nu}
+g_{\mu\nu}(R^{\alpha}_{\phantom{\alpha}\alpha})^2/2$$
$$W^{(2)}_{\mu\nu}
=g_{\mu\nu}(R^{\alpha}_
{\phantom{\alpha}\alpha})
^{;\beta}
_{\phantom{;\beta};\beta}/2  +
R_{\mu\nu\phantom{;\beta};\beta}^{\phantom{\mu\nu};\beta}
 - R_{\mu\phantom{\beta};\nu;\beta}^{\phantom{\mu}\beta}
-R_{\nu\phantom{\beta};\mu;\beta}^{\phantom{\nu}\beta}
 - 2R_{\mu\beta}R_{\nu}^{\phantom{\nu}\beta}
+g_{\mu\nu}R_{\alpha\beta}R^{\alpha\beta}/2
\eqno(9)$$
so that the theory is a lot less tractable than
the Einstein one.
Despite the severe computational
difficulties that the theory possesses, Mannheim and Kazanas$^{2}$
have actually been able to find its complete and exact
solution exterior to a static, spherically symmetric, gravitating
source, viz.
$$-g_{00}= 1/g_{rr}=1-\beta(2-3 \beta \gamma )/r - 3 \beta
\gamma                + \gamma r - kr^2
\eqno(10)$$
where the parameters $\beta$, $\gamma$
and $k$ are
three dimensionful integration constants which appear in the solution but not
in the
equations of motion and thus serve to spontaneously break the scale symmetry.
As we see, the
solution  turns out to be none other than the analog and extension of the
exterior
Schwarzschild vacuum solution of Einstein gravity. We find that in the fourth
order theory the
Schwarzschild solution still obtains except that it is augmented by a new
confining-type
gravitational potential term which grows linearly with distance, which can have
a  strength
such that this new term is unimportant on solar distance scales (so that the
successes of
Einstein gravity on those distance scales remain intact), and which can then
first become
important only galactically. Since the gravitational potential of the theory
grows with
distance it then follows that the rotational velocities of stars in galaxies
should not fall
as a function of distance; and we have recently$^{7}$ obtained some  good
first fitting to the rotation curves of four typical galaxies (each one being a
representative of four characteristic categories of rotation curve which
correlate velocity
with luminosity for the relevant galaxies) in the conformal theory which we
present here as
Fig. (1). As we can see, the rotation  data are able to tolerate the presence
of a linearly
rising potential and admit of reasonable fitting. (The fitting so far has only
been done by
integrating the linear and Newtonian terms over the observed luminous matter
distributions
assuming constant mass to light ratios.
As the microlensing data become more precise it will of course become
necessary to include
such sources in the fitting as well). Intriguingly, we find from the fits that
for a galaxy
the coefficient of its linear term is typically of order the inverse of the
Hubble
radius, to thus suggest an intriguing cosmological (Machian?) connection.  Thus
the viewpoint
of conformal gravity is that the theory of gravity needs to be revised on
galactic and larger
distance scales,  and that since the new potential term does grow with
distance, the
deviations from Newton-Einstein should be even more pronounced on even larger
distance
scales, this also being in accord with the gross observational trend actually
found on
large scales. (Actually treating the theory on large distance scales is quite
subtle because at some point the linear potentials of all of the rest of
the galaxies in the universe become competitive with that of a given galaxy of
interest, with
the given galaxy seeing not the total effect of all the other galaxies - that
only contributes
to the overall Hubble flow - but rather the deviation from homogeneity, a
deviation which
necessitates developing a theory of galaxy formation in the conformal
theory).

As regards
the cosmological implications of the fourth order theory, so far one exact
Robertson-Walker
type solution has been found in a simple scalar field model,$^{5}$ which yields
a
topologically open $k<0$ universe which expands but then nonetheless
recollapses (because of the infrared slavery  associated with the linear
confining
potential), and then rebounds from a finite minimum radius  to oscillate
indefinitely. The model thus provides a closed form exact solution to a  fully
relativistic cosmology which possesses no spacetime singularity at all.
Additionally, the
cosmology possesses no flatness problem (the flatness problem is not a generic
property of
Robertson-Walker cosmologies, it is a specific  feature of the Einstein
Equations, and thus
avoided in the conformal theory which simply has a different set of equations
of motion).
Since there is no  flatness problem in the model, there is thus no need to
appeal to the
popular inflationary universe, so that its need for large amounts of
cosmological dark matter
is thereby avoided.

In the interior region inside of a spherically symmetric, static gravitational
source we find$^{8}$ that in the conformal theory
the fourth order Poisson equation of Eq. (5)
emerges as an exact all order classical equation where the parameter $B(r)$ is
now given
as the metric component  $-g_{00}(r)$ and
where the spherically symmetric source function is given as
$$f(r) = {3 (T^0_{{\phantom 0} 0} - T^r_{{\phantom r} r})/4
\alpha B(r)}   \eqno(11)$$
Recognizing that the radial piece of $\nabla^4 B(r)$ can be written as
$(rB)^{\prime \prime \prime \prime}/r$, we thus immediately see that our
exterior metric of Eq. (10) emerges as the most general solution to
the fourth order Laplace equation $\nabla^4 B(r)=0$. Moreover, we can also
match
the solution of Eq. (6) to that of Eq. (10) to enable us
to express the parameters of the exterior solution
in terms of moments of the interior matter
distribution according to
$$\beta(2-3\beta \gamma)
= {1 \over 6} \int_{0}^{R} dr^{\prime} f(r^{\prime}) {r^{\prime}}^
4~~~,~~~\gamma = -{1 \over 2} \int_{0}^{R} dr^{\prime} f(r^{\prime})
{r^{\prime}}^2 \eqno(12)$$
\noindent
We thus see the somewhat unanticipated outcome that even though
the Green's function $-\vert {\bf r}- {\bf r^{\prime}} \vert /8 \pi$ of
the fourth order $\nabla ^4 $ operator is linear in the distance $r$, after
integrating Eq. (5) with it to actually obtain Eq. (6), we find that we
obtain not merely the linear
potential term, but also the $1/r$ Newtonian term of Eq. (6) as well,
even though no second order
Laplacian operator $\nabla ^2 $ is present anywhere in the fourth
order theory. Thus we see that while a second order Poisson equation is
sufficient to generate a $1/r$ potential, it is not in fact necessary,
with Newton's Law obtaining in the fourth order theory also.

As a potential macroscopic gravitational theory, conformal gravity admits of a
wealth
of eventual astrophysical and cosmological testing; and, indeed, it is
paramount to study the
dynamics of clusters of galaxies in the theory (where the effects of the linear
potential will
be even more prominent than in the individual galaxies themselves) as well as
the detailed
structure of galactic gravitational lensing (the most
sensitive consequence of dark matter); while on stellar distance scales
it is important to study the decay of the
orbit of a binary pulsar in the theory (a test of the field nature of gravity
and of the existence of gravitational radiation reaction which is present
in any relativistically covariant theory where gravitational information
is communicated retardedly with a finite velocity).
On cosmological distance scales it is necessary to study the
implications of the theory as they impact on the isotropy of the cosmic
microwave background, on the cosmological mass content of the universe,
on the development of large scale
structure, on the growth of inhomogeneities and galaxy formation, and
on primordial nucleosynthesis. (The initial analyses (L. Knox and  A. Kosowsky,
Fermilab preprint Pub-93/322-A, and also D. Elizondo and G. Yepes, Ap. J. 428,
17
(1994)) of
nucleosynthesis in the single scalar field model cosmology presented in Ref.
(5) is that the
theory  produces sufficient primordial helium but apparently not enough of the
other light
elements. It is
thus crucial to develop a full theory of galaxy formation to see whether the
associated presence of inhomogeneities can improve the situation).
All of these galactic and cosmological astrophysical phenomena
can ultimately provide for definitive observational testing of
both conformal gravity and the standard second order Einstein theory.
\medskip
\noindent
{\bf (3) Conformal Gravity as a Microscopic Theory}
\medskip
While the solution of Eq. (12) shows that the Newtonian term will be
obtainable for any spherically symmetric matter distribution, we note
that its strength is related to the fourth moment of $f(r)$ rather than
to the second one, the case which occurs in the familiar second order
Einstein theory. Since this fourth moment would vanish
for a delta function source, we see that in order to obtain a Newtonian
potential in the fourth order theory the source must be extended rather
than pointlike. While this violates our standard second order intuition
(but not any observational information incidentally - the $1/r$ potential is
the exact
exterior solution to the second order Poisson equation in Eq. (4) no matter how
the source
$g(r)$ behaves in the interior region, with a delta function source being
sufficient
but not at all necessary to yield the Newtonian solution in the exterior), it
is not all
that surprising since our experience with dynamical mass generation in the
other
fundamental interactions (which we recall motivated our choice of locally
conformal invariant gravity in the first place) indicates that we should
anyway expect elementary particles to be
extended soliton or bag-like objects rather than pointlike ones, with the
only new feature here being that curvature must now play a role in producing
such
structures.

To emphasize this point, consider scalar and spinor fields $\psi(x)$
and $S(x)$  coupled conformally to gravity with matter action
$$I_M=-\int
d^4x(-g)^{1/2}[S^\mu S_\mu/2+\lambda S^4-S^2R^\mu_{\phantom
{\mu}\mu}/12+i\bar{\psi}\gamma^{\mu}(x)(\partial_\mu+\Gamma_\mu(x)) \psi
-hS\bar{\psi}\psi] \eqno(13)$$
where $\Gamma_\mu(x)$ is the fermion spin connection and $h$ and $\lambda$ are
dimensionless coupling constants. As we can see, when the Ricci scalar is
non-zero, it can
induce a tachyonic Higgs mass term into a theory with no fundamental scales and
no
fundamental $-{\mu}^2S^2$ term at all. Thus we propose that even though the
curvature outside
of elementary particles may be small, nonetheless curvature effects within them
may well be
substantial (perhaps even induced by gluon exchange, which itself may even be
another way of describing curvature effects simply because of the equivalence
principle). With this curvature effect then giving the
fermion a Higgs mass in Eq. (13), we may then find that such elementary
particles only have small curvature effects on each other; i.e. that just
like in the nuclear shell model, most of the interaction is used up in
producing the self-consistent states in the first place, with these
states then only interacting with each other through weak residual forces.
In this way microscopic curvature could be very strong, with it then
almost all being used up just in order to produce massive fermions in the first
place,
with these fermions then only having
weak residual gravity, so that we have to go to large macroscopic
systems before the effects can build up enough to become
observable. To establish such a picture we must thus look for extended bag
like fermionic solutions based on Eq. (13) and its analogs.

Since the solutions in our theory are extended according to the above
discussion, there is some concern as to
whether such extended objects are compatible with
positivity of the source distribution $f(r)$. To this end we have
constructed an explicit though only illustrative candidate source which
is positive definite everywhere, and which may be thought to describe
extended elementary particles. Its specific form
is
$$f(r) =-2 p
\delta(r)/r^2 - (3q/2) [\nabla^2 - (r^2/12) \nabla^4][\delta(r)/r^2]
\eqno(14)$$
where $p$
and $q$ are new intrinsic parameters which characterize the fundamental source,
and its
positivity (for an appropriate choice of the signs of $p$ and $q$) may be made
manifest by writing the $q$-dependent part of the source as the limit $\epsilon
\rightarrow 0$ of
$$f(r)=6q
\epsilon (9r^4-3 \epsilon ^4 -10 \epsilon ^2 r^2) / \pi (r^2 + \epsilon ^2 )^5
\eqno(15)$$
with $f(r)$
then being positive in this limit while trapping a singularity at the
origin. For the source function  of Eq. (14) we find that Eq. (12) yields $
\beta (2-3\beta
\gamma )=q$ and $\gamma = p$, which shows that, because of the specific nature
of the
trapped singularity, the strengths of the
linear and Newtonian potential terms of a fundamental source are in principle
independent,
with there thus being no relation of the form $\gamma \sim \beta/R^2_0$ where
$R_0$ is the radius of the source. Thus, as might perhaps have been expected in
a higher
derivative theory, an elementary particle now comes with two rather than one
intrinsic mass
scales,  a intriguing result which requires further
study. Since the weak gravity potential outside of a fundamental source is
just given by $ V(r)=-q/2r+pr/2$, we see that for macroscopic weak gravity bulk
matter of mass
$M$, the coefficient of its ensuing macroscopic $1/r$
potential term is given as $Nq/2$, i.e. directly  proportional to the total
number, $N$, of
microscopic singularities just as in the Einstein case where the same
coefficient
is identified as $MG$. Thus the
very deep singularity in Eq. (14) takes care of the microscopic fourth moment
integral and
leads us to a macroscopic inverse square gravitational force when $r \ll
1/\gamma$ which is
universal (because $\alpha$ is universally coupled in Eq. (7)), which is an
extensive function of the number of particles (as long as binding effects
are unimportant), and which for spherical bulk matter only depends on the
distance from  the center of the source, just as desired for Newton's Law. Thus
having
a fundamental Newton constant is only sufficient to give universal gravitation,
but not
apparently necessary. ($G$ is never defined independently in Newtonian gravity,
only the
product $MG$ is ever measured gravitationally - Newton's constant thus has a
status not
unlike that of Boltzmann's constant which only ever appears in the product
$kT$).
In the above way we can thus build up the macroscopic potentials of stars from
their
fundamental constituents and then use the stars as sources to calculate the
weak gravity
galactic geodesics which are used to fit the rotation curves shown in Fig. (1).
(Of course, simply because of their spherical symmetry, stars will have
potentials of the
form $V(r)=-\beta/r+\gamma r/2$ (numerically the $\beta \gamma$ product in the
$1/r$ term in
Eq. (10) is found to be negligible) because that is the exact potential in Eq.
(10)
no matter how these stellar coefficients depend on the microscopic
substructure. Thus
the phenomenological fitting of Fig. (1) is completely insensitive to specific
details of
the fundamental microscopic $f(r)$). While the source
function given in Eq. (14) is not of course mandated as the only  possible
source appropriate
to the theory, its very existence serves as a counter-example to the claim
often made in the
literature  that in the fourth order theory Newton's law is incompatible with
positivity of the matter distribution, and thus enables us to dispose of an
objection to the conformal theory which had previously hindered
its development.

As a microscopic theory the conformal theory suffers from two well known
difficulties,
namely the theory is found to possess ghosts and anomalies in lowest order
perturbation
theory when quantized around flat spacetime. As regards the trace anomaly, we
note, that
unlike its triangle anomaly counterpart, the trace anomaly is renormalizable.
Thus as well as
possibly being canceled by a group theoretical interplay between the
fundamental fields of
the theory (as part of a unification?) or by a non-trivial interplay between
graviton and
matter field loops (or even by quantizing around some other geometric
background), it is also
in principle possible to cancel it non-perturbatively by a Gell-Mann Low
coupling
constant renormalization eigenvalue  (see e.g. S. L. Adler, J. C. Collins, and
A. Duncan,
Phys. Rev. {\bf D15}, 1712 (1977)) which would then restore the  underlying
scale invariance
of the theory only with anomalous dimensions. (With the advent of asymptotic
freedom,
renormalization group fixed points away from the origin fell somewhat into
disfavor.
Nonetheless, they still represent a viable non-perturbative option for field
theory which has
never been formally excluded; and it was noted quite some time ago (P. D.
Mannheim, Phys.
Rev. {\bf D12}, 1772 (1975)) that when the dimension of the fermion
$\bar{\psi}\psi$ bilinear is
reduced from 3 to 2 by a fixed point in QED, the vacuum then undergoes
spontaneous breakdown
and generates dynamical masses. It would thus be of some
interest to see if an analogous situation obtains in the conformal case). As
regards
linearizing around flat spacetime in general, we note that unlike Einstein
gravity where the
matter free background is Riemann flat Minkowski space, for the conformal
theory the natural
background is a vanishing Weyl tensor, with the background metric then not
being flat  but
only conformal to flat. (From the point of view of maximally symmetric 4
spaces, de Sitter,
anti de Sitter and Minkowski spaces all have the maximal 10 Killing vectors,
and all are flat
or conformal to flat. Thus the background for gravity should perhaps be
maximally symmetric
rather than flat, with explicit energy-momentum tensor sources to which gravity
couples then
lowering the symmetry in specific physically interesting cases).  Further,
given the
non-asymptotically flat metric of Eq. (10), the flat spacetime limit may not be
all that
relevant to the theory, with the restoring linear potential possibly even
confining the
ghosts and removing them from the physical spectrum in the true vacuum
altogether (perhaps
with a Gell-Mann Low eigenvalue actually setting the ghost residue to zero just
like the
analogous Landau ghost cancellation in quantum electrodynamics - moreover, if
the
dimension of $g_{\mu \nu}$ could be changed by one whole unit at such an
eigenvalue, the
$1/q^4$ propagator could then be non-perturbatively modified into the ghost
free $1/q^2$).
As regards the ghost question, we note additionally that in a first order
perturbation
expansion around flat spacetime the fourth order theory yields a fourth order
box operator,
to thus yield corrections to the metric which grow with distance (just like the
linear
potential) and which hence become indefinitely large at large distances. Since
the  ghost
states would appear on shell at low momenta, we see that their presence in the
theory is
inferred in precisely the kinematic region where first order perturbation
theory becomes
untrustworthy. (Nonetheless, the ghosts are still free to contribute at very
short distances
far off the mass shell, to thus maintain the power counting  renormalizability
that the
fourth order theory is known to have - basically the $1/q^4$ propagator is
equivalent to two
$1/q^2$ propagators, one being a regular graviton, and the other a ghost
graviton which
cancels the familiar non-renormalizable infinities associated with a single
Einstein
propagator). Thus, if anything, the ghosts are signaling only that flat
spacetime is not a
good limit to the theory, something which would anyway be expected given the
structure of
the exact, non-perturbative solution of Eq. (10). It is thus both worthwhile
and crucial to
explore the ghost and anomaly issues further, and in particular to quantize the
theory around
the metric of Eq. (10) rather than around flat
spacetime.

As a microscopic theory the
conformal theory also has many positive features. It is known to be
renormalizable
(K. S. Stelle, Phys. Rev. {\bf D16}, 953 (1977)), asymptotically free (E.
Tomboulis,
Phys. Letts. {\bf 97B}, 77 (1980)),
and as we have now seen, explicitly confining. In that sense then it begins to
compete with
QCD. Moreover, given the analog between local gauge invariance and local scale
invariance
(the gauge transformation is a complex phase on the fields while the scale
transformation is
real one), and given the intriguing analog between the conformal
$C_{\lambda\mu\nu\kappa}C^{\lambda\mu \nu\kappa}$ and gauge $F_{\mu\nu}F^{\mu
\nu}$ actions,
local scale invariance might provide the   road to unification of gravity with
the other
interactions by making gravity look a lot more  like the others. While work
still needs to be
done on the ghost and anomaly questions, we  recall that string theory, another
candidate
quantum gravitational theory, languished for many years until M. B. Green and
J. H. Schwarz
(Phys. Letts. {\bf 149B}, 117 (1984))
found a very clever higher dimensional anomaly cancellation mechanism.
Certainly  the enormous
effort that has been made on string theory by a whole generation of researchers
has yet to be
made on the conformal theory, and it too could possess surprises; indeed  much
of our
current understanding of strings and of gauge  theories and  of particle
physics in general
grew out of a need to cancel ghosts, so a ghost often proves to be a highly
informative
diagnostic. Thus at the very least conformal gravity merits further study.
Since conformal
gravity sets out to be a completely consistent quantum theory of gravity
(perhaps via a
Gell-Mann Low eigenvalue), it sets out to provide an alternative to string
theory, and,
moreover, to do so directly in four  spacetime dimensions from the beginning.
In fact the
coupling constant $\alpha$ in the  conformal action of Eq. (7) is only
dimensionless in four
spacetime dimensions.  Moreover, since conformal gravity comes with no
fundamental scale (it
demotes Newton's  constant from fundamental status), it stands in sharp
contrast to string
theory which gives extra special status to Newton's constant over all other
dimensionful
fundamental physical  constants. It may, however, be possible to establish some
connection
between the conformal  and string  theories in the zero string tension limit
where string
theory then possesses no intrinsic  scale. In fact if a connection could be
made, then as
long as the known consistency of  string theory survives in this limit, we
would then be able
to infer the consistency of  scaleless gravity too and thus provide for a
string based
mechanism for solving the  conformal theory ghost problem non-perturbatively.
As regards the
general issue of the  relative merits of string gravity  and conformal gravity,
we note that
one of the nice features of having a conventional renormalizable gravitational
theory such as
conformal gravity is that, just like QED, there should then be a direct
correspondence
between the microscopic and macroscopic theories, with the matrix elements of
the quantum
field in  coherent states of its quanta then having a chance to have equations
of motion of
the same generic form as those of the underlying quantum field itself, simply
because the
primary role of renormalization is to then convert bare propagators and
vertices into dressed
ones. In this way the action of Eq. (7) could then be used for both microscopic
and
macroscopic physics.  To contrast, we note that in string theory, however, the
connection between microscopic and macroscopic fields is somewhat remote.
Further, we note
that string theory is not so much an attempt  to construct a consistent quantum
gravitational
theory, but rather to construct one which reduces to Einstein gravity at low
energies. (It
does not quite do that incidentally since  while it nicely recovers the
Einstein term it also
leads to a huge Planck density cosmological  term as well. String theory thus
makes the
cosmological constant problem more not less  severe, and this might even be an
indicator of
its lack of viability). As we have now seen  in our discussion of macroscopic
gravity, there
may not in fact be any need to actually recover the  Einstein Equations anyway,
with the
central question for microscopic physics in the end  actually being what
exactly the
non-relativistic gravitational potential is  on the largest distance scales.
\medskip
\noindent
{\bf (4) Conformal Gravity and the Structure of the Energy-Momentum Tensor}
\medskip
As we noted earlier, it was only with the emergence of spontaneously
broken mass generation in strong, electromagnetic and weak interactions,
that it became possible to return to the scaleless conformal gravitational
theory. However, it is quite remarkable that the general community has shown no
interest in
incorporating any of this new Higgs based physics into the standard
gravitational Einstein
theory. Indeed, one of the most curious and disquieting aspects of the
entire contemporary approach to the role of mass in fundamental theory is that
mass is treated as being of dynamical origin for the purposes of the
strong, electromagnetic and weak interactions, and yet is treated as being
purely mechanical and kinematical for gravitational
purposes; with the source of Einstein gravity being taken to be not of the
Weinberg-Salam or QCD form, but rather to simply be a collection of mechanical
particles with their familiar kinematic energy-momentum tensors, i.e. to be of
the form
thought to prevail when General Relativity itself was first written down. Thus,
at the
present time, essentially all macroscopic tests of General Relativity simply
assume that
gravity is produced by perfect fluid sources which are purely kinematic in
structure
possessing purely mechanical energy densities and pressures. This purely
Newtonian
description of motion (after correction for relativistic kinematics) forms the
cornerstone of
gravitational  exploration, and is regarded as being so sacrosanct and obvious
as to not even
require any further justification, even though such sources have never been
shown to emerge
in gauge theories, and even though string theory gives little guidance as to
the explicit
structure of the right hand side of Eq. (2).

That this Newtonian model for $T_{\mu \nu}$  was actually considered at all is
because of too
much of a reliance on the flat space limit of gravity, a limit in which one
only measures the
mechanical motions of particles as they move about, i.e. in which one only
measures energy and
momentum changes. In flat spacetime the zero of energy is not observable, only
the energy of the particle excitations with respect to the vacuum, and yet the
wisdom
obtained from these excitations is then written in a general covariant form and
posited as the source of gravity without further question. Now gravity responds
not merely to energy and momentum changes, but also to the zero of energy and
momentum, something which is of course recognized as a part of the
cosmological constant problem and then ignored. Perhaps even more serious than
not knowing how to set the zero of energy is the fact that we now know that
particles get their masses (which then characterize the one particle
excitations via the energy-momentum relation $E^2_k=k^2+m^2$) from Higgs fields
which also carry energy and momentum, with this energy and momentum also being
ignored in the standard treatment, again without justification. Thus in
order to correctly include all of the Higgs effects and determine the zero of
energy consistently, we not only need a theory of mass generation in a theory
with no
intrinsic scales, we also
need to incorporate its implications into the structure of the gravitational
source.

Contemporary particle physics leads us quite naturally to matter actions such
as the scaleless Higgs one exhibited in Eq. (13), and even if one does not want
to make the jump to conformal gravity on the gravitational side of the
gravitational equations of motion, one should thus at least use the
energy-momentum
tensor associated with Eq. (13), viz.$^6$
$$T_{\mu \nu} = i \bar{\psi} \gamma_{\mu}(x)[ \partial_{\nu}
+\Gamma_\nu(x)]
\psi+2S_\mu S_\nu /3 -g_{\mu\nu}S^\alpha S_\alpha/6 -SS_{\mu;\nu}/3
+g_{\mu\nu}SS^\alpha_{\phantom{\alpha};\alpha}/3$$
$$-S^2(R_{\mu\nu}-g_{\mu\nu}R^\alpha_{\phantom{\alpha}\alpha}/2)/6
-g_{\mu\nu}\lambda S^4 \eqno(16)$$
\noindent
as the source on the right hand side of the Einstein Equations in Eq. (2) since
Eq. (16)
and its analogs embody much of the wisdom of particle physics. Use of the
scalar field
equation of motion
$$S^\mu
_{\phantom{\mu};\mu}+SR^\mu_{\phantom{\mu}\mu}/6-4\lambda S^3+h\bar{\psi}\psi=0
\eqno(17)$$
and of the covariant Dirac equation
$$i \gamma^{\mu}(x)[\partial_{\mu} +\Gamma_\mu(x)]
\psi - h S \psi = 0 \eqno(18)$$
enable us to show that the conformal energy-momentum tensor is both
kinematically traceless
and covariantly conserved just as it should be. From Eq. (16) it is
possible$^{6,8}$ to
derive some crucial results for gravitational theory. First, as can be seen
directly from
Eq. (16), in the constant $S(x)=S_0$ gauge not only is the full $T_{\mu \nu}$
covariantly conserved, but two separate pieces of it, namely the kinematic
fermion
piece $T_{\mu \nu}^{kin} = i \bar{\psi} \gamma_{\mu}(x)
[\partial_{\nu}+\Gamma_\nu(x)]\psi$
and the remaining Higgs dependent piece $T_{\mu \nu}(S_0)$, are independently
covariantly
conserved also. Thus, while the fermion could share energy and momentum with
the Higgs field
which gives it its mass, it turns out that in fact it does not. Hence the one
particle
excitations of the fermion are the same as they would be if the fermion simply
had a
mechanical bare mass, with matrix elements of the fermionic piece of the
energy-momentum
tensor being the same as the $T_{\mu \nu}^{kin}$ conventionally used in the
standard theory,
so that its associated covariant conservation then gives geodesic motion for
such particles
in an external field. Second, an incoherent averaging over a bath of the
fermions of Eq. (18)
then gives the fermion kinetic term the form of a kinematic perfect fluid, viz.
$T^{kin}_{\mu\nu}=(\rho+p)U_\mu U_\nu+pg_{\mu\nu}$ where $\rho= \int dk k^2
E_k/ \pi^2$,  $p=
\int dk k^4/ 3 \pi^2 E_k$,  with its covariant conservation then giving$^{6,8}$
the familiar
Euler hydrostatic equation when the fermions are the source of the
gravitational field. Hence
we see that both geodesic motion and perfect fluid hydrodynamics still occur in
theories with
mass generation, and that in and of themselves, they are simply totally
misleading guides as
to the structure of the full energy-momentum tensor of gravitational theory,
since even while
the particle motions may decouple from the Higgs fields, the gravitational
field itself is
still very much sensitive to them. Since this latter Higgs energy and the
attendant back
reaction on the geometry  contained in $T_{\mu \nu}(S_0)$ are simply
ignored in the standard Newtonian based picture of gravitational sources
(even though $T^{\alpha}_{\phantom {\alpha} \alpha}(S_0)$
is equal to $-T^{\alpha}_{\phantom {\alpha} \alpha}(kin)$ and thus as large),
at the
present time the entire standard treatment of Eq. (2) must be  regarded as
suspect. Thus for
instance, if the source of gravity is the traceless Eq. (16), it would then
lead in the
standard theory to $R^{\alpha}_{\phantom {\alpha} \alpha}=0$ and thus to
cosmologies in which
the scale factor
$R(t)$ grows as
$t^{1/2}$ even in the matter dominated era ($\rho=nm$), thus making it quite
unclear as to
what the standard theory would then produce by way of nucleosynthesis. Thus, at
the
present time, the nucleosynthesis successes of the standard theory must be
regarded purely as a coincidence until some explanation is found for why we can
model the entire $T_{\mu\nu}$ in Eq. (2) simply as $T^{kin}_{\mu\nu}$. And,
moreover, the ignoring of the back reaction of the Higgs field on the geometry
is in fact the
cosmological constant problem.

If the reader is now prepared to make the full conformal jump on the
gravitational side as well we are thus led to replace the Einstein Equations
altogether by Eq. (8) coupled to Eq. (16) and its analogs, with a symmetry,
conformal
invariance, then dictating the structure of both sides of the gravitational
equations, something which has up till now not been achieved in the standard
theory. Since the entire theory is now prescribed, the cosmological term has no
freedom, with there being no kinematic one in the equations of motion, and with
the back reaction on the geometry sharply constraining its value if one is
induced dynamically, with the tracelessness constraint actually forcing its
magnitude to be the same as that of the contribution of the positive energy
matter fields
which propagate cosmologically$^{1,5}$ and
not 120 orders of magnitude larger (not that the conformal theory even
possesses a
Planck scale anyway). However, to solve the cosmological constant problem
completely we still need to consider the contribution of the negative energy
modes which fill the vacuum. These, of course, are the modes which give masses
to the fundamental particles in the first place, with their back reaction on
the geometry causing particles to become extended in the first place, so again
there are constraints. In fact, based on Eqs. (8), (16-18), the following
model of elementary particles is suggested. These
equations are found to admit of an exact solution in which a
kinematic perfect fluid of fermions couples to a constant scalar field in a
static Robertson-Walker geometry with spatial curvature $k$. In the solution
the energy
density is found to be given by $\rho=-\lambda S^4_0-S^2_0k/2$, so that the
negative energy
modes can thus support a closed $k>0$ universe, with elementary particles then
emerging as
trapped three surfaces (black holes are only trapped two surfaces). (In passing
we note
that
$\rho >0$ essentially gives the $k<0$ cosmology of Ref. (5)). Since classically
no energy or
momentum, or even the vacuum energy, can be transported across trapped three
surfaces, such
classical systems may not suffer from the classical radiation reaction problem.
Quantum
mechanically they might also be closed and thus confining, though they may
still be able to
communicate with other particles by Hawking-like radiation. In this way then
the  negative
vacuum energy of a filled Dirac sea could be nicely trapped inside of
elementary particles
and never be manifest on cosmological distance scales. (Within such elementary
particles the
curvature based string picture may still be able to emerge, though possibly
with the old dual
resonance model hadronic string tension instead). As we can thus see, particle
physics has
some interesting implications for the structure of gravitational sources, with
the
cosmological constant problem perhaps being not so much a problem for
fundamental theory in
general, but  possibly only being a problem for theories based on the use of
the
Einstein-Hilbert action; with the problem then being not so much how to get rid
of the
cosmological constant at all, but rather how to get rid of it in an
Newton-Einstein based
theory where it has no reason not to be there.

Given that the Newton-Einstein theory was first established on solar distance
scales in the
weak gravity limit, the theoretically so far unjustified
insistence on the use of the Einstein Equations for both larger distances and
stronger
coupling thus represents a quite enormous extrapolation to kinematic regimes
where there
is currently little observational guidance. Indeed, the Einstein theory insists
that
gravity be attractive on all scales, whilst in the conformal theory the back
reaction
$-S^2R^\mu_{\phantom{\mu}\mu}/12$ term in Eq. (13) acts as an induced repulsive
gravitational
term (it has the opposite sign to the Einstein-Hilbert term) which prevents the
cosmology
of Ref. (5) from actually having any singularity.
(That singularities may not be generic to gravity has also been emphasized by
N. J. Cornish and J. W. Moffat in their recent UTPT-94-08 study of the
non-symmetric
gravitational alternative). The full perturbative gravitational expansion for
the metric
could thus look very different from that expected in the Einstein theory, and
only
observations on large distances and in strong fields can ultimately enable us
to identify the
rest of the series.  This work has been supported in part by the Department of
Energy under
grant No. DE-FG02-92ER40716.00.
\medskip
\baselineskip=0.40truecm
\noindent {\bf References}
\medskip
\item{1.}
         P. D. Mannheim, {\it Gen. Rel. Grav.} {\bf  22}, 289 (1990).

\item{2.}
         P. D. Mannheim and D. Kazanas, {\it Ap. J.} {\bf 342}, 635 (1989).

\item{3.}
         D. Kazanas and P. D. Mannheim,  {\it Ap. J. Suppl. Ser.} {\bf 76}, 431
(1991).

\item{4.}
         P. D. Mannheim and D. Kazanas, {\it Phys. Rev.} {\bf D44}, 417 (1991).

\item{5.}
         P. D. Mannheim, {\it Ap. J.}  {\bf 391} 429 (1992).

\item{6.}
         P. D. Mannheim, {\it Gen. Rel. Grav.} {\bf 25}, 697 (1993).

\item{7.}
         P. D. Mannheim, {\it Ap. J.}  {\bf 419}, 150 (1993).

\item{8.}
         P. D. Mannheim and D. Kazanas, {\it Gen. Rel. Grav.} {\bf 26}, 337
(1994).

\item{9.}
         P. D. Mannheim,  {\it Founds. Phys.} {\bf 24}, 487 (1994).
\medskip
\baselineskip=0.44truecm
\noindent
{\bf Figure Caption}
\medskip
\noindent
Figure (1). The calculated rotational velocity curves of Ref. (7)
associated with stellar
potentials $V(r)=-\beta/r+\gamma r/2$ for four typical galaxies,
the intermediate sized NGC3198, the compact luminous NGC2903,
the large sized NGC5907, and the dwarf irregular DDO154 (which is calculated at
both of
two possible observational distances from the Milky Way). In
each graph the bars show the observed data points with their quoted errors as a
function of
radial distance (in arc minutes) from the center of each galaxy. The full
curve shows the overall theoretical galactic rotational velocity prediction (in
km/s),
while the two indicated dotted curves show the rotation curves that the
separate stellar
Newtonian and linear potentials would produce when integrated over the luminous
matter distribution of each galaxy. No dark matter is assumed.
\end